\documentclass[aps,prl,preprint,superscriptaddress,showpacs]{revtex4-1}

\usepackage[version=3]{mhchem} 
\usepackage{graphicx}
\usepackage{color}
\usepackage{soul}

\newcommand{\bq}{\mathbf{q}}
\newcommand{\br}{\mathbf{r}}
\newcommand{\bG}[1][]{\mathbf{G}_{\mathrm{#1}}}

\usepackage{amsmath} 

\begin{document}

\title{Quantum plasmons with optical-range frequencies 
in doped few-layer graphene}

\author{Sharmila N. Shirodkar}
\thanks{Present address: Department of Materials Science and NanoEngineering, Rice University, Houston, TX 77005, USA}
\email{sns8@rice.edu}

\author{Marios Mattheakis}
\altaffiliation{Department of Physics, University of Crete, PO, Box 2208, 71003 Heraklion, Greece}
\affiliation{John A. Paulson School of Engineering and Applied Sciences, Harvard University, Cambridge, Massachusetts 02138, USA}
\author{Paul Cazeaux}
\affiliation{School of Mathematics, University of Minnesota, Minneapolis, Minnesota 55455, USA}
\author{Prineha Narang}
\altaffiliation{Faculty of Arts and Sciences, Harvard University, Cambridge MA 02138, USA}
\affiliation{John A. Paulson School of Engineering and Applied Sciences, Harvard University, Cambridge, Massachusetts 02138, USA}
\author{Marin Solja\v{c}i\'{c}}
\affiliation{Department of Physics, Massachusetts Institute of Technology, 77 Massachusetts Avenue, Cambridge, Massachusetts 02139, USA}
\author{Efthimios Kaxiras}
\affiliation{John A. Paulson School of Engineering and Applied Sciences, Harvard University, Cambridge, Massachusetts 02138, USA}
\altaffiliation{Department of Physics, Harvard University, Cambridge, Massachusetts 02138, USA}
\date{\today}

\begin{abstract}
Although plasmon modes exist in doped graphene, the limited range of 
doping achieved by gating restricts the plasmon frequencies to a range that 
does not include visible and infrared.  Here we show,  
through the use of first-principles calculations, 
that the high levels of doping achieved by 
lithium intercalation in bilayer and trilayer graphene 
shift the plasmon frequencies into the visible range.
To obtain physically meaningful results,
we introduce a correction of the effect of plasmon interaction across the vacuum 
separating periodic images of the doped graphene layers, 
consisting of transparent boundary conditions in the direction perpendicular to the layers;
this represents 
a significant improvement over the Exact Coulomb cutoff technique employed in earlier works.
The resulting plasmon modes are due to local field efffects and the non-local 
response of the material to external electromagnetic fields, requiring a fully quantum 
mechanical treatment. 
We describe the features of these quantum plasmons, 
including the dispersion relation, losses and field localization.
Our findings point to a strategy for fine-tuning the plasmon 
frequencies in graphene and other two dimensional materials.
\end{abstract}

\maketitle 

Collective excitations of electrons in metals, generically referred to as plasmons, 
have been attracting new attention recently 
in the realm of nanoparticles and low-dimensional materials.  
In these systems, new plasmonic phenomena continue to be discovered,
beyond what was observed in conventional crystalline solids.  These phenomena  
include quantum interference of plasmons, 
observation of quantum coupling of plasmons to single particle excitations,
and quantum confinement of plasmons in nm-scale particles and materials.
These phenomena, intriguing in their own right, are also important for multifaceted 
applications. 
Plasmonic nanostructures are finding applications in integrated nanophotonics \cite{schuller2010}, biosensing \cite{bender_2009,friedt_2004,rodrigo_2015},
photovoltaic devices \cite{atwater2010, schaadt2005,clavero2014}, single photon transistors \cite{chang2007},
single molecule spectroscopy \cite{zhang_2013} and metamaterials \cite{liu_2013,marios2016}.
The current interest in quantum nanophotonics and plasmonics is in part driven by new materials,
particularly low dimensional solids, that access new ranges of frequency and transmission 
speeds.
The reduced dimensionality of plasmons in two-dimensional (2D) materials provides ultra-subwavelength confinement with phase velocities
several orders of magnitude lower than the speed of light \cite{andress_2012}.
In the present work we show that by properly controlling the density of metallic electrons 
in few-layer graphene, the prototypical 2D metal, the plasmon 
frequency can be pushed into the visible to near-infrared range, 
a feature highly desirable for 
optoelectronic applications and heretofore unattainable.  

Graphene is quite special for 2D plasmonics \cite{basov2012}, exhibiting 
intriguing properties such as extremely high electrical mobility \cite{novoselov2004} and 
easily tunable electron and hole doping concentrations ($n_e, n_h$), through gating \cite{novoselov2004,waghmare2008}.
The plasmon frequencies in graphene are controlled through doping~\cite{basov2012},
where typical doping concentration values achieved 
by gating are $\approx 10^{11}$ cm$^{-2}$,
and the heaviest doping reached~\cite{yan_2013} is $n_h > 10^{13}$ cm$^{-2}$.
Plasmons in gate-doped graphene typically emerge in the infrared to THz ranges, 
and seldom in the mid- or near-infrared range \cite{rodrigo_2015,yan_2013,novoselov_2012}.
So far, reaching the visible range for 2D plasmons in graphene,
a crucial requirement for optoelectronic applications, has remained elusive. 
Searching for materials beyond graphene to achieve plasmons with optical frequencies is 
a possible route.  For example, one possibility is the family 
of 2D materials referred to as transition metal dichalcogenides (TMDCs),
but plasmons in these materials 
are predicted to appear at THz frequencies \cite{thygsen_2013,thygsen_2015}.
Another possible solution, 
the plasmon mode on Be(0001) \cite{basov2012} observed in the visible range \cite{bogdan_2007}, cannot be interpreted as a true 2D plasmon, 
since it has finite penetration depth into the underlying bulk material.
A recent report by Huang {\it et al.} \cite{huang_2017} predicts that triangular polymorph of 2D boron sheet exhibits visible frequency plasmons.
But free-standing triangular 2D boron is dynamically unstable \cite{alex_2016} and its experimental synthesis quite difficult, which makes it challenging for device applications. 

We propose here an alternative approach for breaking the impasse, 
by doping few-layer graphene structures to levels beyond what is achievable through gating.
Though there have been previous reports of optical-frequency plasmons in
graphene monolayers with adsorbed Li atoms (LiC$_2$)~\cite{despoja2017}, 
this configuration is energetically unstable as we have established in previous work~\cite{sharmila_2016}, and therefore unlikely to form experimentally; 
encapsulating the Li atoms between graphene layers,
as in the structures proposed and studied here, is required to stabilize the doped system.
Experiments have proved the feasibility of inserting metal atoms like lithium (Li)
between layers of 2D materials \cite{guo_2016,sugawara_2011} resulting in heavy doping. 
Inspired by this, 
we use a theoretical approach based on first-principles electronic structure calculations 
to explore the possibility of observing quantum plasmons in the visible range 
for Li-intercalated two- and three-layer graphene.
The origin of 2D plasmons is related to the local field effects and the non-local response of the material to external fields \cite{raza_2011}. 
Hence, the study of these waves demands a fully quantum mechanical 
description of the material properties 
which compells us to call them as \textquoteleft quantum\textquoteright~2D plasmons.
We effectively capture the quantum nature of these plasmons through our accurate,
high-fidelity first-principles calculations, distinguished by: 
(i) our methodology which correctly confines plasmons in two dimensions, and (ii) a
realistic estimate of carrier lifetime, a crucial factor that determines plasmon losses.
Our results show that quantum plasmons in few-layer graphene are indeed feasible.
This opens new pathways for fine-tuning a wide range of plasmon frequencies,
including the visible range, in 2D structures, by 
controlling the concentration and type of intercalants.


Our first-principles calculations are based on density functional theory (DFT)
as implemented in the GPAW package \cite{gpaw1,gpaw2}.
The interaction between ionic cores and valence electrons is described by
the projector augmented wave method \cite{paw1,paw2}.
A vacuum of 25 \AA~is included to minimize the interaction between periodic images
along the direction perpendicular to the plane of the sheets ($z$ direction).
The Kohn-Sham wavefunctions are respresented using a plane wave basis with energy cutoff of 340 eV,
and the exchange correlation energy of electrons is described using Local Density Approximated (LDA) functional.
For the linear response calculations, used to estimate the dielectric functions \cite{linear_response}, we sample the Brillouin zone
with a 256 $\times$~256 $\times$~1 grid of k-points to include an accurate description of intraband transitions.
For the dielectric response calculations we use a plane wave energy cutoff of 30 eV.
All the other parameters were converged to within 0.05 eV of the plasmon energies, 
using the methodology developed by Andersen {\it et al.}~\cite{thygsen_2012, thygsen_2013} for calculating the quantum plasmon modes.

The potential $\phi(\br, \omega)$ and charge density $\rho(\br, \omega)$
of the quantum plasmon modes, are obtained as left and right eigenfunctions (which satisfy the Poisson equation)
of the dielectric operator $\hat{\epsilon}(\omega)$, diagonalized in the plane wave basis:
\begin{equation}\label{eq1}
	\hat{\epsilon}(\omega)\phi_n(\omega) 
= [ \hat{1} - \hat{v}\  \hat{\chi}^0(\omega) ] \phi_n(\omega) 
= \lambda_n(\omega) \phi_n(\omega),
\end{equation}
where $\omega$ and $\br$ denote the frequency and in-plane spatial vector, respectively.
Here, $\hat{\epsilon}(\omega)$ expressed in terms of the noninteracting linear response 
operator $\hat{\chi}^0(\omega)$ and the Coulomb interaction operator $\hat{v} = 1/\vert \br - \br' \vert$. 
The condition for observing a plasmon at frequency $\omega_\text{p}$ is $\mathrm{Re}[\lambda_n(\omega)] =0$ 
or equivalently a peak in the loss function, $-\mathrm{Im}[\lambda_n(\omega)]/|\lambda_n(\omega)|^2$.

A key ingredient in obtaining the plasmon dispersion relations
and losses is the carrier lifetime, $\tau$. 
To obtain reliable values of $\tau$, we used DFT results 
for the energies and matrix elements of both electrons and phonons (see Supplemental Material \cite{supp} and \cite{prineha_2017adv}).
This takes into account the detailed electronic structure effects 
such as response of electrons far from the Dirac point,
as well as scattering against both accoustic and optical phonons including Umklapp and inter-valley processes
\cite{prineha_2016acs, prineha_2017adv,brown_2016prb,brown_2016experimental}.
Doping, that is, change in position of the Fermi level ($E_\text{F}$), changes the value 
of $\tau$,
and hence calculations were carried out for several different values of $E_\text{F}$ ranging from the neutral (undoped) value to 1.5 eV above it
(see Supplemental Material \cite{supp} for details of formulation and \cite{prineha_2017adv} for values of $\tau$).
Interestingly, our results show that the extremely 
large $\tau\approx$ 1 ps for free standing undoped graphene
drops to $\approx$ 29 fs in doped graphene.
For simplicity and computational efficiency, we use a doped  monolayer graphene
to obtain the values of $\tau$ for positions of $E_\text{F}$ that correspond 
to those of the Li-doped bilayer and trilayer graphene; this is a reasonable approximation,  
because, at high doping concentrations, we expect that
the effects of interlayer electron-phonon and electron-electron 
coupling on $\tau$ in intercalated graphene 
will be rather small compared to the effects of 
changing the position of E$_F$, which is properly taken into account
by the procedure described.

The standard approach for eliminating spurious effects due to finite size of vacuum \cite{rozzi_2006} is inadequate
for plasmons with small in-plane wavectors ($\bq$), and increasing the size of the vacuum region until these effects become negligibly small 
requires very expensive calculations.
A significant methodological contribution of the present work is the formulation and implementation of transparent boundary conditions
which overcome the drawbacks of the Coulomb cutoff method and offer a more accurate description of the quantum plasmon fields.
Let $z_-$, $z_+$ be the bounds of the super-cell (simulation box) along the $z$ direction (vacuum region) with $(x,y)$ plane being periodic. 
We apply a one-dimensional Fourier transform in the $z$ direction to obtain 
a real space representation in this coordinate. The response operator under random phase approximation (RPA)  then has the form:
\begin{equation}
\label{eq3}
	 \hat{\chi}^0 \phi (z, \bG[xy], \bq, \omega) =\\ \int_{z_-}^{z_+} \sum_{\bG[xy]'} \chi^0_{\bG[xy],\bG[xy]'}(z,z',\bq,\omega) \phi(z', \bG[xy]', \bq, \omega) \mathrm{d}z',
\end{equation}
where $\bG[xy]$, $\bG[xy]'$ are vectors of the in-plane reciprocal lattice
For values of $z,z'$ inside the super-cell,
$z_- < z,z' < z_+$, the kernel $\chi^0_{\bG[xy],\bG[xy]'}(z,z')$ 
is deduced from $\chi^0_{\bG,\bG'}$ by Fourier transform.
The kernel is extended with zero values
for $z$ or $z'$ that lie in the vacuum region outside this cell.
We observe that Eq.~\eqref{eq1} can be reformulated 
as the generalized eigenvalue problem \cite{supp}:
\begin{equation}
\label{eq4}
	\hat{\chi}^0 \phi_n(z, \bG[xy], \bq, \omega) =\\ \frac{1 - \lambda_n}{4\pi} \left (\vert \bq + \bG[xy] \vert^2 -  \frac{\partial^2}{\partial z^2}\right ) \phi_n(z, \bG[xy], \bq, \omega),
\end{equation}
with additional constraint that $\vert \phi_n\vert  \to 0$ as $z \to \pm \infty$ so the problem is well-posed.
The left-hand side vanishes in the vacuum region and Eq.~\eqref{eq4} reduces to the one-dimensional Poisson equation.
For any nonzero value of $\vert \bq + \bG[xy]\vert$, we thus obtain an explicit solution 
\begin{equation*}
\phi_n(z, \bG[xy], \bq, \omega) = \phi_n(z_\pm, \bG[xy], \bq, \omega) e^{ - \vert \bq + \bG[xy] \vert \vert z_\pm- z\vert },
\end{equation*}
$\text{for } z \leq z_-~\text{and}~z \geq z_+$. The continuity of $\phi_n$ and its first derivative with respect to $z$ leads to the  transparent boundary conditions at $z = z_\pm$:
\begin{equation}\label{eq5}
\frac{\partial \phi_n}{\partial z}(\bq, \bG[xy], z_\pm, \omega) = \mp \vert \bq + \bG[xy] \vert \phi_n(\bq, \bG[xy], z_\pm,\omega),
\end{equation}
which implies that the charge density and potential do not see the periodic boundary along
the $z$ direction for any value of $\bq$, and hence decay to zero as $ z \rightarrow \pm \infty$.
The imposition of additional constraints generalizes the previous approaches \cite{rozzi_2006,despoja_2013}, which makes the transparent boundary conditions
an improvement over the former techniques.
We solve numerically by finite differences the eigenvalue problem of Eq.~\eqref{eq4} restricted to the finite band $z_- \leq z \leq z_+$,
with the boundary conditions of Eq.\eqref{eq5} (see Supplemental Material for details \cite{supp}).

\begin{figure}[!ht]
\includegraphics[width=16cm]{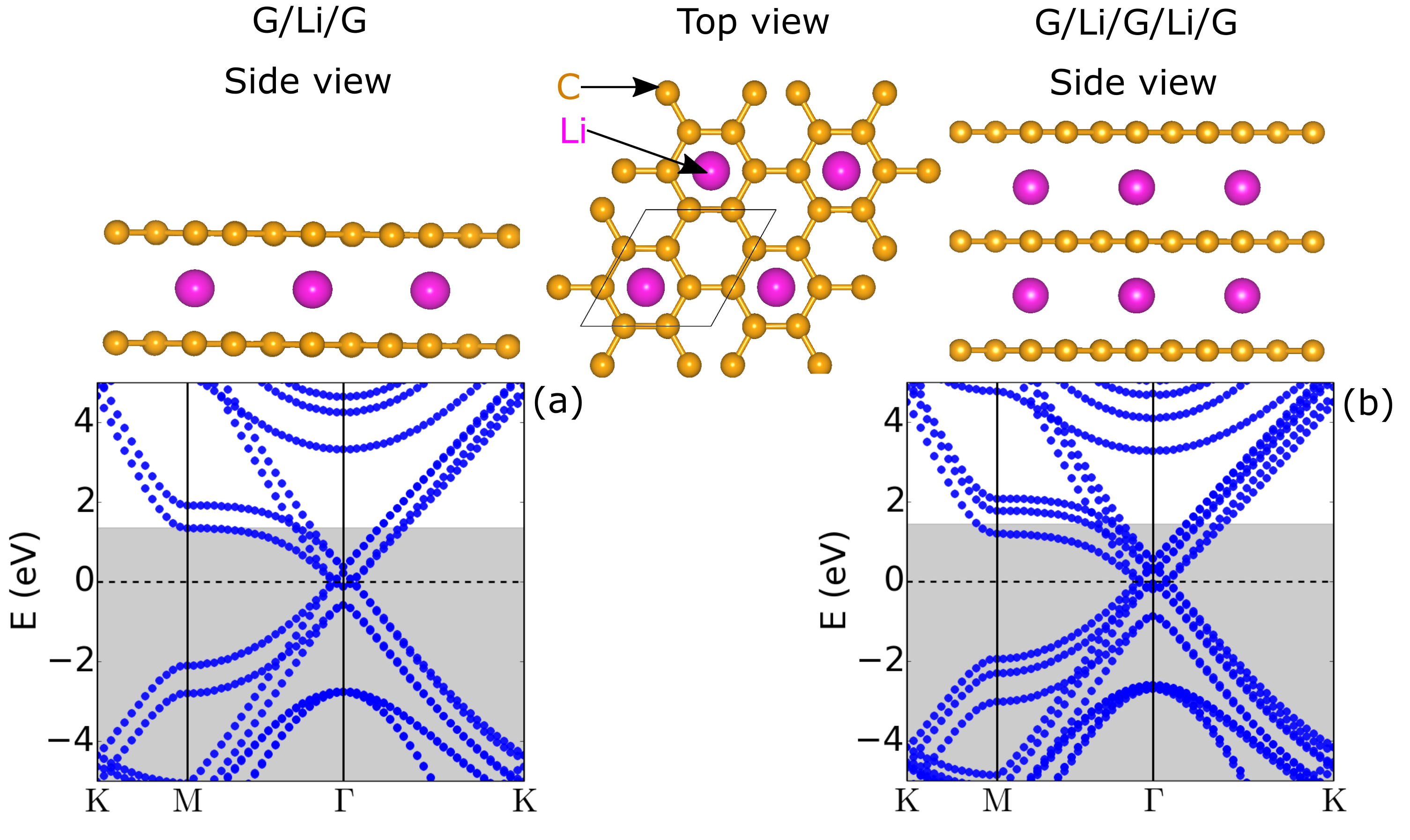}
\caption{Atomic structures (side and top views) and electronic structures 
of: (a) the bilayer Li-intercalated graphene (G/Li/G, left) and 
(b) the trilayer Li-intercalated graphene (G/Li/G/Li/G, right).
The shaded regions in (a) and (b) denote the occupied states, 
and the dashed black lines the Dirac point / Fermi level in undoped layers.}
\label{fig1}
\end{figure}

We model Li intercalated graphene (G) multilayers with an in-plane
$\sqrt{3}\times\sqrt{3}$ multiple of the primitive unit cell of graphene, with the G/Li/G (bilayer) and G/Li/G/Li/G (trilayer) structures.
There is one Li atom per unit cell between each pair of layers (see Fig. \ref{fig1}) \cite{sharmila_2016, guzman_2014}.
For the trilayer, we consider the structure with the two Li atoms at the same hollow site but between two different pairs of graphene layers,
as this is the most stable configuration \cite{guzman_2014}.
Li intercalation makes AA stacking energetically more preferable \cite{sharmila_2016} and hence both bilayer and trilayer structures are inversion symmetric.
The separation between the layers increases by 0.14~\AA~and 0.11~\AA~relative to its value in the AA stacked graphene bilayer (3.52 \AA), for the bilayer and trilayer, respectively.
Due to band folding in the $\sqrt{3} \times \sqrt{3}$ unit cell,
the high symmetry K point and hence the Dirac point of primitive graphene cell folds onto $\Gamma$ point in the Brillouin zone (BZ) in our simulations (see Fig. \ref{fig1}).
AA stacking preserves the sublattice symmetry of the layers
and the linear dispersion of the electron bands at the Dirac point, unlike AB stacking where the bands are parabolic \cite{low_prl2014}.
Intercalation also leads to charge transfer from Li to the graphene layers, 
and renders the system metallic (see Fig. \ref{fig1}) with $\approx$ 0.84$e$ and 0.87$e$ charge transferred from Li to bilayer and trilayer graphene (determined using Bader analysis),
which corresponds to $n= 5 \times  10^{14}$ and $n= 10^{15}$, respectively. Subsequently, shifting the Fermi level from the Dirac point 
into the conduction band by 1.35 eV  and by 1.51 eV for the bilayer and trilayer, respectively,  as seen in Fig. \ref{fig1}.

\begin{figure}[!ht]
\includegraphics[width=6.5in]{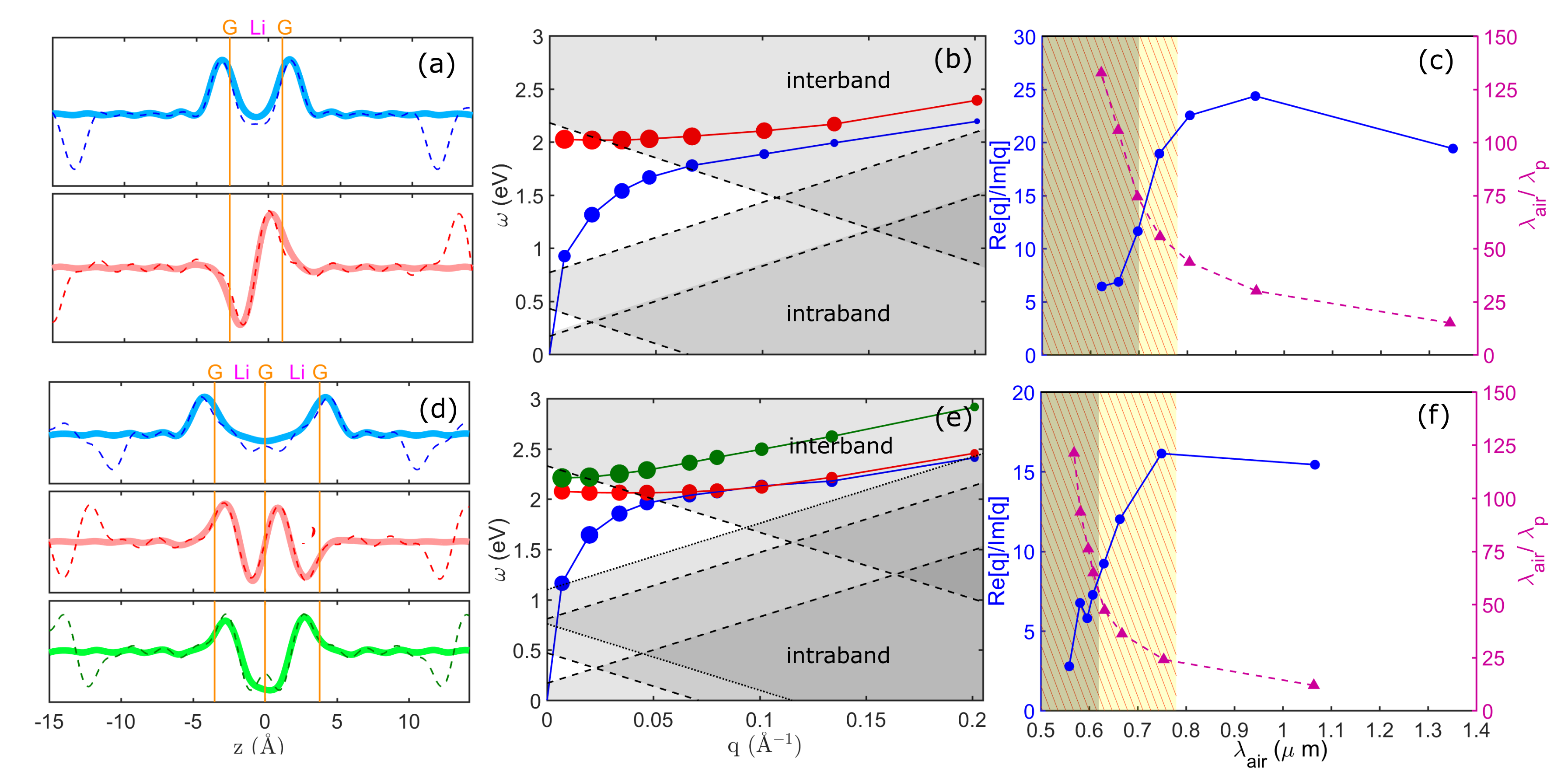}
\caption{Plasmon features for: (a)-(c), the G/Li/G system,  and (d)-(f), the G/Li/G/Li/G system.
(a) and (d) Plasmon charge density $\rho({\bf r})$ at $q$= 0.007 \AA$^{-1}$ for 
the symmetric modes (blue and green lines) 
and the antisymmetric mode (red lines);
solid lines (thicker and lighter shade) are for results with transparent boundary conditions, dashed lines (thinner and darker shade)
for periodic boundary conditions with Coulomb cutoff (see text).
(b) and (e) Dispersion relation of plasmons along the $\Gamma$ to M direction;
the diameter of the circles is proportional to the strength of the resonance \cite{thygsen_2013}.
Shaded areas represent regions of inter- and intra-band losses (including damping by optical phonon).
(c) and (f) Re[$q$]/Im[$q$] (left axis, solid line in blue),
and field localization (right axis, dashed line in magenta), 
or \textquotedblleft shrinkage\textquotedblright, of the lowest symmetric mode.
$\tau$ is $\approx$ 29 fs and 19 fs for the G/Li/G and the G/Li/G/Li/G systems, respectively.
The grey shaded areas denote the region of inter-band losses, 
and the yellow shaded (hatched) areas denote the visible frequency range, 
calculated with the Fermi velocity of graphene.}
\label{fig2}
\end{figure}

Since we consider metallic multilayers, more than one plasmon modes emerge \cite{thygsen_2012, thygsen_2013,low_prl2014}. 
Depending on the phase of the charge density and potential fields, we differentiate them as symmetric and antisymmetric plasmonic modes [see Fig. \ref{fig2}(a) and (d)].
For small $\bq$, the decay length of 2D plasmons extends beyond the vacuum region giving rise to interactions with periodic images,  and hence, spurious fields and pseudo charges at the vacuum edge. 
On the other hand, our transparent boundary conditions correct these periodic interactions and make the plasmon tails invisible to one another for the same vacuum length.
The charge density with (solid lines) and without (dotted lines) transparent boundary conditions is shown in Fig. \ref{fig2}(a) and (d) for G/Li/G and G/Li/G/Li/G, respectively.
We also note that the charge transferred from Li is equally distributed in the unoccupied $\pi^*$ orbitals, which is confirmed from the charge density distribution
of the plasmon modes [see in Fig. \ref{fig2}(a) and (d)], 
where the intensity of the fields is equal and reaches the maximum/minimum values away from the layers,
consistent with the fact that the $\pi^*$ orbitals of graphene extend away from the layers.

We plot the plasmon dispersion along $\Gamma$-M (the $\Gamma$-K direction is not as interesting in the band structure)
with the magnitude of the real part of \textbf{q} ranging from $\vert$\textbf{q}$\vert$= $q$= 0.007 \AA$^{-1}$ to 0.21 \AA$^{-1}$,
since both plasmon modes become very weak above $q$= 0.21 \AA$^{-1}$.
The symmetric mode is more dispersive than antisymmetric, and varies as  $\sqrt{q}$ at small $q$,
corresponding to classical plasmon with Drude behavior due to intraband transitions.
Whereas the antisymmetric mode varies almost linearly with $q$ (has finite frequency at $q$=0)
and relates to interband transitions between perfectly nested bands of the two layers \cite{low_prl2014}.
In G/Li/G the plasmon frequencies are between 0.8 eV to 2.2 eV for $q\geq$ 0.007 \AA$^{-1}$;
the antisymmetric mode is in the optical frequency range even at low $q$, whereas the symmetric mode enters into this range at higher $q$ values.
The symmetric mode is always lower in energy than the antisymmetric mode due to finite coupling~\cite{low_prl2014}.
We note that the acoustic plasmon arising from the anisotropy of the bands crossing the Fermi level along $\Gamma$-M
is not captured in our calculations 
due to limitations of the frequency grid which is too coarse on the 
scale required to reveal this feature. 
However, this does not affect our conclusions since this particular mode is damped by the intraband transitions and therefore not of interest here.

We quantify the plasmon losses from the ratio of real to imaginary component of wavenumber, Re[$q$]/Im[$q$] \cite{jablan_2009},
which corresponds to the number of plasmon wavelengths that propagate before it loses most of its energy [see Fig. \ref{fig2}(c)].
For the doping in G/Li/G ($E_\text{F}$ = 1.35 eV), a $\tau \approx$ 29 fs 
was calculated using our methodology discussed above, which is much shorter in comparison with $\tau \approx$ 135 fs for $E_\text{F}$ = 0.135 eV \cite{jablan_2009} .
We only give the ratio for the symmetric (intraband) mode in Fig. \ref{fig2}(c).
Due to its linear dispersion, antisymmetric mode shows less variation in Re[$q$]/Im[$q$] as compared with symmetric mode (see Supplemental Fig. S1 \cite{supp}).
The in-plane propagation length of the plasmons varies directly with this ratio, with the symmetric plasmons propagating longer at longer wavelengths ($\lambda_{\text{air}}$).
We also calculate the wave \textquotedblleft shrinkage\textquotedblright or the field localization of the plasmons, shown in Fig. \ref{fig2}(c);
this corresponds to the ratio by which the plasmon wavelength ($\lambda_\text{p}$) is smaller than that in vacuum, and is approximately 100 times for bilayer graphene.

There are three important decay modes that lead to plasmon damping:
(i) Landau damping due to intraband losses when $\hbar\omega$\textless~ $\hbar v_{\text{F}}q$, 
(ii) interband losses (electron-hole transitions
referred to as single-particle excitations, 
SPE's, identified as poles of the response function \cite{low_prl2014,jain_1987}) 
when $\hbar\omega$\textgreater~$\hbar\omega_{\text{SPE}}$ (with damping region defined by $\hbar\omega_{\text{SPE}} - \hbar v_{\text{F}}q$ \textless$\hbar\omega$ \textless$\hbar \omega_{\text{SPE}} + \hbar v_{\text{F}}q$), and 
(iii) decay through optical phonons in graphene for $\omega$\textgreater~$\omega_{\text{ph}}$ ($\omega_{\text{ph}}$= 0.2 eV or 6.2 $\mu$m) \cite{jablan_2009} due to scattering of electrons (that is, plasmonic excitation) due to phonons. 
This calculation of dielectric function under the Random Phase Approximation (RPA)
does not include the effects of electron-hole interactions, which are captured only by including a dynamically screened instead of the bare Coulomb interaction.
However, these excitons give rise to a prominent peak in the absorption spectrum near 4.5 eV \cite{yang_2011} which is at a much higher energy than the visible frequency range.
Also, doping has been shown to increase screening and reduce electron-hole interactions in graphene, leaving the optical response nearly identical to undoped graphene \cite{yang_2011}.
Hence, the exclusion of electron-hole interactions in our calculations does not affect the results.

In case of G/Li/G, since the optical phonon $\omega_{\text{ph}}$ = 1400 cm$^{-1} \equiv$ 0.17 eV \cite{guzman_2014,profeta_2012} is
much smaller than the symmetric or antisymmetric plasmon frequencies (0.8 eV to 2.2 eV for $q\geq$~0.007 \AA$^{-1}$),  
only multiple scatterings by phonons (which are less likely) will scatter plasmons into the damping regions.
On the other hand, plasmons within frequency range $\omega_{\text{SPE}} - \omega_{\text{ph}}$ to $\omega_{\text{SPE}}$ can get scattered by phonons into Landau/interband scattering regions.
Therefore making $\omega$\textgreater$~\omega_{\text{SPE}}-\omega_{\text{ph}}$ the region where plasmons are damped by interband transitions and optical phonons.
The SPE's at $q$= 0 were identified at 0 eV, 0.6 eV and 2.4 eV originating from the intraband,
low energy interband and the electron-hole interband transitions in G/Li/G.
The damping regions are defined by E$_{\text{SPE}}\pm \hbar v_{\text{F}}q  ~\pm~\hbar\omega_{\text{ph}}$~(including scattering by optical phonons), 
where $v_{\text{F}}$ is the Fermi velocity and E$_{\text{SPE}}$ is the single particle excitation energy \cite{brey_2013,jain_1987} [see gray shaded areas in Fig. \ref{fig2}(b)].
Heavy doping by lithium pushes the electron-hole interband threshold for the bilayer to $\omega_{\text{inter}}\approx 1.77$ eV ($\lambda = 0.7  \; \mu$m).
Since the optical frequency range ($\omega_{\text{op}}$) is between 1.59 eV to 3.26 eV 
($\lambda = 0.38 \; \mu$m to 0.78 $\mu$m) and 
$\omega_{\text{inter}}<\omega_{\text{op}}$,
most of the symmetric and antisymmetric plasmon modes in this range are not damped 
by the interband transitions, indicated by the shaded regions in Fig. \ref{fig2}(b) and \ref{fig2}(e).
Only for $q\geq 0.06$ \AA$^{-1}$ are the symmetric and antisymmetric modes damped.

To push the interband threshold frequency, and hence the plasmon frequencies, 
higher into the optical range ($> 2$ eV), the Fermi level needs to be moved farther into 
the conduction bands.
Since the maximum possible intercalation in bilayer graphene 
corresponds to composition C$_{12}$Li, 
additional Li can be incorporated only by having more than two graphene layers.
We therefore explore trilayer graphene since it can accomodate two Li layers, 
with a composition Li$_2$C$_{18}$,
which increases the doping level to $E_\text{F}$ = 1.51 eV.
There are three modes in the trilayer structure in the 1.2 -- 2.8 eV frequency range 
along the $\Gamma$-M direction for $q\geq$~0.007 \AA$^{-1}$,
two of which are symmetric and one antisymmetric, shown in Fig. \ref{fig2}(d).
The third (second symmetric) mode emerges due to the third graphene layer which brings in additional nesting of the bands.
Similar to the bilayer case, the first symmetric mode due to intraband excitations exhibits $\sqrt{q}$ dependence and the other two modes disperse linearly, see Fig. \ref{fig2}(e).
The loss function shows larger variations in the peak positions for the first symmetric mode due to $\sqrt{q}$ behavior at low $q$
as compared to the antisymmetric mode (see Supplemental Fig. S2  for details \cite{supp}). 
More interestingly, the first symmetric and antisymmetric bands in the dispersion 
spectrum [red and blue curves in Fig. \ref{fig2}(e)] intersect
and the symmetric and antisymmetric modes are degenerate for $q >0.067$ \AA$^{-1}$ along $\Gamma$-M.
The reason behind this unusual degeneracy is the \st{fine} nesting between the bands 
at the Fermi level and consequently the absence of coupling between the two modes \cite{low_prl2014}.

The higher doping concentration pushes the interband threshold frequency ($\omega_{\text{inter}}$) to $\approx$ 2.0 eV (0.62 $\mu$m) for 
the first symmetric and antisymmetric modes in G/Li/G/Li/G.
The poles at 0 eV, 0.64 eV, 0.93 eV and 2.5 eV correspond 
to the three damping regions associated with intraband, low energy interband, and
higher energy electron-hole interband transitions.
Hence, for 1.59 eV $<\omega<$ 2.0 eV
(0.62 $\mu$m$<\lambda_{\text{air}}<0.78 \; \mu$m)
the first symmetric and antisymmetric modes are undamped. 
More importantly, the second symmetric mode gets damped at a higher frequency ($\omega > 2.2$ eV),
so all three plasmon modes are undamped and emerge in the optical range for $q< 0.05$ \AA$^{-1}$.
The $\tau$ in graphene for such high doping concentration ($E_\text{F}$ = 1.51 eV) 
is quite small $\approx$ 19 fs (See Supplemental Material \cite{supp}).
From the Re[$q$]/Im[$q$] in Fig. \ref{fig2}(f), we find that the first symmetric mode can be observed 
further into the mid-infrared range (from extrapolation)
($\lambda_{\text{air}}>3 \; \mu$m), whereas the other two modes have shorter 
wavelengths ($\lambda_{\text{air}}<0.62 \; \mu$m).
$\lambda_\text{p}$ is also shrunk by approximately 100 times, Fig. \ref{fig2}(f),
as in the case for bilayer graphene, in agreement with previous reports \cite{jablan_2009}.
We only plot the ratio for the first symmetric (intraband) mode in Fig. \ref{fig2}(f).
Since the antisymmetric and second symmetric modes disperse linearly, 
the variation in the Re[$q$]/Im[$q$] is small.
These plasmons exhibit similar \textquotedblleft shrinkage\textquotedblright as that of 
the symmetric mode (refer to Supplemental Fig. S2 for further details \cite{supp}).

Controlling the number of layers and the concentration of intercalated Li atoms appears 
to be a feasible method for engineering the properties of visible plasmons for applications.
For example, the mid-infrared region plasmons in both the bilayer and 
trilayer Li-intercalated structures, 
can be used for plasmonic biosensing \cite{rodrigo_2015,yan_2013}.
We caution that certain technical aspects of the calculations reported here, 
like the choice of exchange correlation functional for the electronic structure, 
can affect the electronic spectrum and can shift the plasmon energies to slightly different 
values than what we reported; such shifts could change the precise values of the 
damped plasmon frequencies but we do not expect them to alter the overall picture.
Damping due to the presence of defects and substrate phonons, features that 
were not included in the model of the physical system considered here,  
can also influence the existence of undamped 2D plasmons in the visible frequency range. 
A detailed analysis of these parameters will constitute the future scope of this work.
Our work can be easily extended to explore other multilayers of other 2D  
materials (such as black phosphorus, transition metal dichalcogenides)
with different dopants and/or intercalants (K, Mg, Na etc), 
opening up new pathways for fine tuning the plasmon dispersion
either by varying the number and type of layers, and/or by varying 
the concentration and type of intercalant atoms.

\begin{acknowledgements}
The authors thank R. Sundararaman and J. Joannopoulous for ab-initio calculations and discussions related to plasmon lifetimes.
The authors also thank J. Cheng, S. Inampudi, H. Mosallaei and G. A. Tritsaris for useful discussions.
MM acknowledges support from EU program H2020-MSCA-RISE-2015-691209-NHQWAVE.
We acknowledge support by ARO MURI Award No. W911NF14-0247 (SNS and EK)
and by EFRI 2-DARE NSF Grant 1542807 (MM).
PN and MS were partly supported by the Army Research Office 
through the Institute for Soldier Nanotechnologies under contract no.
W911NF-13-D-0001.
We used computational resources on the odyssey cluster of the Research Computing Group at Harvard University,
and at the Extreme Science and Engineering Discovery Environment (XSEDE), which is supported by NSF Grant No. ACI-1053575.
\end{acknowledgements}

\clearpage
\section{Figures}

\clearpage

\bibliography{ref}

\end{document}